# Moiré Magnetic Exchange Interactions in Twisted Magnets

Baishun Yang[1,2], Yang Li[1], Hongjun Xiang[3], Haiqing Lin[4,1,5], and Bing Huang[1,5]

[1]*Beijing Computational Science Research Center, Beijing 100193, China*
[2]*Shenzhen JL Computational Science and Applied Research Institute, Shenzhen 518109, China*
[3]*Key Laboratory of Computational Physical Sciences (Ministry of Education), Institute of Computational Physical Sciences, State Key Laboratory of Surface Physics, and Department of Physics, Fudan University, Shanghai 200433, China*
[4]*School of Physics, Zhejiang University, Hangzhou 310058, China*
[5]*Department of Physics, Beijing Normal University, Beijing 100875, China*

Correspondence should be addressed to Bing Huang at Bing.Huang@csrc.ac.cn.

**Besides moiré superlattice, twisting can also generate moiré magnetic exchange interactions (MMEIs) in van der Waals magnets. However, due to the extreme complexity and twist-angle-dependent sensitivity, all existing models fail to capture the MMEIs, preventing the understanding of MMEIs-induced new physics. Here, we develop a microscopic moiré spin Hamiltonian that enables the effective description of MMEIs via a sliding-mapping approach in twisted magnets, as demonstrated in twisted bilayer $CrI_3$. Unexpectedly, we discover that the emergence of MMEIs can create an unprecedented magnetic skyrmion bubble (SkB) with non-conversed helicity, named as moiré-type SkB, representing a unique spin texture solely generated by MMEIs and ready to be detected under the current experimental conditions. Importantly, the size and population of SkBs can be finely controlled by twist angle, a key step for skyrmion-based information storage. Furthermore, we reveal that the MMEIs can be effectively manipulated by the substrate-induced interfacial Dzyaloshinskii-Moriya interaction, modulating the twist-angle-dependent magnetic phase diagram, which solves the outstanding disagreements between prior theories and experiments.**

**Main text**

Moiré superlattices, which serves as ideal platforms to host fascinating properties and create intriguing applications, can be constructed by stacking two van der Waals (vdW) layered materials with a relatively small twist angle ($\theta$) [1,2]. Such a twist angle offers a new degree of freedom to effectively modulate the fundamental electronic structures, providing an exotic approach to generate and manipulate many new physical phenomena in different 2D vdW systems, e.g., moiré flat bands [3], unconventional superconductivity [4-7], moiré Hubbard and Kane-Mele-Hubbard models [8,9], moiré charge orders [10], anomalous Hall effect [11], and moiré excitons [12-15]. Importantly, developing effective theories to capture the electronic structures of extreme large-scale moiré lattice play an indispensable role in understanding these interesting moiré-type phenomena[3,4,9,15]. These exciting advances are introducing a brand-new research area in condensed matter physics, i.e., twistronics [16,17].

Recently, many efforts are made to bring the vdW magnets into twistronics, to generate exotic spin textures and unconventional magnetic phase transitions (MPTs) [18-30]. Differing from the conventional magnets with a few simple magnetic exchange interactions, the moiré lattice in twisted magnets could induce large-scale periodic magnetic exchange interactions with more than tens of thousands of nonequivalent magnetic parameters, forming complicated moiré magnetic exchange interactions (MMEIs). "More is different", as stated by Philip Anderson in 1972 [31], indicating the uniqueness of MMEIs in generating new magnetic phenomena. Unfortunately, due to the extreme complexity and $\theta$-sensitivity, MMEIs in twisted magnets have never been clearly understood. As a result, many puzzling experimental observations on the complex spin textures and MPTs in twisted magnets cannot be understood. The bilayer $CrI_3$ (BL-$CrI_3$) is such an excellent example [32-34]. In twisted BL-$CrI_3$ (tBL-$CrI_3$) with small $\theta$, the unexpected spin textures with nonzero remanent magnetization can appear forming hexagonal patterns[18-21], raising puzzles on its origin. Unfortunately, all the existing theories [22-30], which are unable to describe the MMEIs, fail to produce these unusual spin textures. Therefore, developing an effective moiré spin Hamiltonian to capture MMEIs is the key step to unveil the mystery of MMEIs in twisted magnets and resolve the outstanding disagreements between existing theories and experiments.

In this article, we have developed a microscopic moiré spin Hamiltonian to accurately describe the MMEIs in twisted magnets via the sliding-mapping approach. Taking tBL-$CrI_3$ as a material example, we demonstrate that the appearance of MMEIs can induce a new type of magnetic skyrmion bubble (SkB) with non-conversed helicity, named as moiré-type SkB. We demonstrate that this unique SkB can solely be generated by the MMEIs in twisted magnets. Interestingly, by finely tuning the small $\theta$, the size and population of these moiré-type SkBs can be precisely controlled, providing a new way for skyrmion-based information storage and quantum computing. Furthermore, we reveal that the substrate-induced interfacial Dzyaloshinskii-Moriya (DM) interaction can largely modulate the intrinsic MMEIs and in turn govern the $\theta$-dependent MPTs, which can perfectly solve the outstanding disagreements between prior theories and experiments and validate our developed theory. Overall, our study provides a benchmark for understanding the MMEIs and MMEIs-induced new physics.

**Results**
**I. An Effective Moiré Spin Hamiltonian in Twisted Magnets.**
Without loss of generality, taking tBL-$CrI_3$ as an example, we develop a unified microscopic spin Hamiltonian that enables the description of MMEIs in twisted magnets. The long-periodic moiré pattern in tBL-CrI$_3$ is formed by twisting one monolayer respected to the other one with a small $\theta$, as shown in **Fig. 1a**. Importantly, the local crystal structures in the moiré pattern are nearly equivalent to a series of BL-CrI$_3$ formed by smoothly sliding the top layer by a vector $\boldsymbol{r}$ = m$\boldsymbol{a_1}$+n$\boldsymbol{a_2}$ with respect to bottom layer from the AA-stacking structure, as shown in **Fig. 1b**, where $\boldsymbol{a}_1$ and $\boldsymbol{a}_2$ are the in-plane unit vectors of the monolayer CrI$_3$. For example, the surrounded insets of **Fig. 1a** show four typical local crystal structures, i.e., AA, AB, BA, and AB', whose interlayer sliding vectors are (m, n) = (0, 0), (2/3, 2/3), (1/3, 1/3), and (0, 1/3), respectively. The general moiré spin Hamiltonian, which can be used to accurately describe the magnetic properties of tBL-CrI$_3$, should include the site-dependent MMEIs as follows:

$$H = \sum_{i,j} J_{inter}^{(i,j)} \boldsymbol{S}_i \cdot \boldsymbol{S}_j + \sum_{i,j} J_{intra}^{(i,j)} \boldsymbol{S}_i \cdot \boldsymbol{S}_j + \sum_{i,j} \boldsymbol{D}_{inter}^{(i,j)} \cdot (\boldsymbol{S}_i \times \boldsymbol{S}_j) + \sum_{i,j} \boldsymbol{D}_{intra}^{(i,j)} \cdot (\boldsymbol{S}_i \times \boldsymbol{S}_j) + K \sum_i (\boldsymbol{S}_i)^2 + H_{ext} \quad (1),$$

where, $\boldsymbol{S}_i$, $\boldsymbol{S}_j$ are spin operators at sites $i, j$, $J_{inter}^{(i,j)}$ and $J_{intra}^{(i,j)}$ represent the interlayer and intralayer Heisenberg exchange couplings, $D_{inter}^{(i,j)}$ and $D_{intra}^{(i,j)}$ represent the interlayer and intralayer DM exchange couplings, respectively. The DM interactions encourage tilting the spins, generating non-collinear spin textures. $K$ indicates the effective strength of the single-ion anisotropy. $H_{ext}$ is the external interaction term induced by external fields or substrate effect. The summations for $i, j$ run over all neighboring Cr atoms within the length scale of BL-CrI$_3$ unit cell.

While directly solve the parameters in **Eq. (1)** is very challenging due to its extreme complexity in a large moiré superlattice, instead, a sliding-mapping approach is developed to indirectly solve these MMEI parameters. **Fig. 1c** shows a typical example to calculate the $J_{inter}$ in tBL-CrI$_3$. The middle panel shows the enlarged local atomic structure of tBL-CrI$_3$ marked as blue frame region in **Fig. 1a**. Here, four typical inter-atomic $J_{inter}$ with different distance $r$ ($r_1$, $r_2$, $r_3$ and $r_4$) are labeled. To obtain the $J_{inter}(r)$ between two Cr atoms in tBL-CrI$_3$: first, we determine the distance $r$ between the two Cr atoms; second, in BL-CrI$_3$, sliding the top layer CrI$_3$ with respect to bottom layer with the vector $r$, we can see the local atomic position in tBL-CrI$_3$ around $J_{inter}(r)$ is nearly the same with that in BL-CrI$_3$ after sliding $r$. Therefore, a direct mapping of the $J_{inter}(r)$ in tBL-CrI$_3$ to $\mathcal{J}_{inter}(r)$ in BL-CrI$_3$ (see **Fig. 1d**) can be generated. In short, as the local structures in tBL-CrI$_3$ are equal to that of BL-CrI$_3$ generated via smooth sliding, the magnetic parameters between two Cr atoms at each local site in the tBL-CrI$_3$ ($D$ and $J$) could be effectively derived through mapping the magnetic parameters of BL-CrI$_3$ ($\mathcal{D}$ and $\mathcal{J}$) by smoothly changing the sliding vector $r$ (see more explanation in **Supplemental Note I**). In principle, as long as the sliding process is sufficiently smooth, one could always get an approximate one-to-one correspondence of the magnetic parameters between BL-CrI$_3$ and tBL-CrI$_3$. In particular, the smaller the $\theta$, the more accurate the mapping parameters one could obtain. Therefore, this approach is very suitable for twisted magnets that usually has very small $\theta$. A similar idea is also used in a very recent moiré excitons simulation[15]. Because it is impossible to consider all the MMEIs under |m, n| > 1 and also because the strength of MMEIs will rapidly

decrease for larger m and n, only the MMEIs within |m, n| ≤ 1 region are considered to ensure the interlayer [nearest to tenth nearest-neighboring ($J_{inter}^{1N}$-$J_{inter}^{10N}$ and $D_{inter}^{1N}$-$D_{inter}^{10N}$)] and intralayer ($J_{intra}^{1N}$-$J_{intra}^{9N}$ and $D_{intra}^{1N}$-$D_{intra}^{10N}$) couplings in a hexagonal lattice are fully included in the moiré superlattice (see Supplemental **Fig. S1**).

The essential difference between **Eq. (1)** and previous models is, for the first time, the complete site-dependent magnetic interactions can be captured between two different magnetic atoms, i.e., the MMEIs could be generated and well described. Most importantly, the relatively weak DM interaction in MMEIs is also considered in a site-dependent way. As shown in the following discussions, such a comprehensive description of MMEIs, especially for the weak DM interactions, are critical to produce the unconventional spin textures in the twisted magnets.

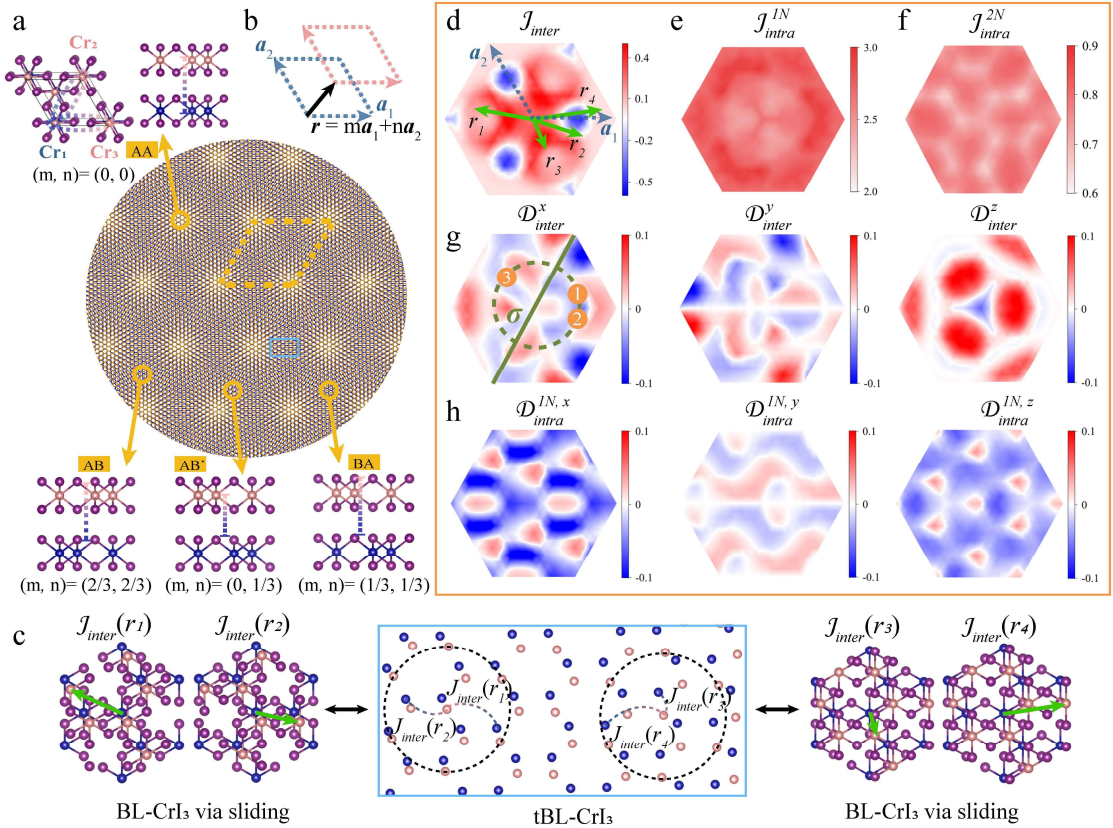

Fig. 1 **Generating MMEIs via Sliding-mapping Approach. a**, Moiré superlattice of the tBL-CrI₃. Surrounded insets show the regions with local AA, AB, AB' and BA stackings, where Cr in the top and bottom layers are colored as the gold and blue, respectively. Dashed-yellow line indicates the unit cell of moiré superlattice. Blue frame represents the local site in Fig. 1**c**. **b**, Schematic diagram of the relative position of the interlayer Cr atoms with the sliding vector **r**. **c**, Examples of one-to-one correspondence between $J_{inter}(r)$ in tBL-CrI₃ to $\mathcal{J}_{inter}(r)$ in BL-CrI₃ via sliding. Color mapping of **d**, interlayer exchange interaction $\mathcal{J}_{inter}$, **e**, intralayer nearest-neighboring exchange interaction $\mathcal{J}_{intra}^{1N}$, **f**, intralayer next-nearest-neighboring exchange interaction $\mathcal{J}_{intra}^{2N}$, **g**, interlayer DM vector $\mathcal{D}_{inter}$, and **h**, intralayer DM vector $\mathcal{D}_{intra}^{1N}$ between two Cr atoms as a function of sliding vector *r*. Here, all the magnetic parameters are in the unit of meV.

**II. Unique Features of Equivalent MMEIs.**

The magnetic interactions in BL-CrI$_3$ with different sliding vectors *r* are plotted in **Fig. 1d-h**. Similar to previous studies, the calculated interlayer exchange $\mathcal{J}_{inter}$ is mostly FM ($\mathcal{J} > 0$) but with three AFM ($\mathcal{J} < 0$) patches within the unit cell (**Fig. 1d**) [35]. However, fundamentally different from previous models [23-26], three important features of $\mathcal{J}_{inter}$ can be observed: (1) The $\mathcal{J}_{inter}$ differs when the top layer slides to (m, n) = (1, 0) and (1, 1). Because the C$_{3v}$ symmetric point group of AA-stacking BL-CrI$_3$ lacks $\sigma_h$ symmetry operation, the exchange path between Cr$_1$ in the bottom layer and Cr$_2$ and Cr$_3$ in the top layer, i.e., Cr$_1$-Cr$_2$ and Cr$_1$-Cr$_3$ pairs (marked in **Fig. 1a**), differs from each other; (2) The $\mathcal{J}_{inter}$ shows $c_3$ but not $c_6$ rotation symmetry due to the same reason; (3) Except the AFM patch regions, the $\mathcal{J}_{inter}$ gets smaller as a function of sliding vector *r*, which is due to the decreased exchange integral as the distance between two interlayer Cr atoms increases. For the intralayer exchange coupling, the relative positions of three nearest and six next-nearest neighboring intralayer Cr-Cr pairs are not changed, but their exchange paths are influenced by the sliding vector *r*. Therefore, the $\mathcal{J}_{intra}^{1N}$ (**Fig. 1e**) and $\mathcal{J}_{intra}^{2N}$ (**Fig. 1f**) are not homogenous anymore, despite without the dramatic changes with the sliding vector *r*. Similarly, $c_3$ rotation symmetry is also observed in the $\mathcal{J}_{intra}^{1N}$ and $\mathcal{J}_{intra}^{2N}$, and the $\mathcal{J}_{intra}^{2N}$ slightly changes compared to that of $\mathcal{J}_{intra}^{1N}$.

The symmetry of AA-stacking BL-CrI$_3$ breaks when the top layer CrI$_3$ slides as a function of *r*, leading to the emergence of DM interaction. Similar to $\mathcal{J}_{inter}$, the interlayer DM $\mathcal{D}_{inter}$ can also be influenced by the interlayer stackings, as shown in **Fig. 1g**. Interestingly, the *x* and *y* components ($\mathcal{D}_{inter}^x$ and $\mathcal{D}_{inter}^y$) show the symmetric and antisymmetric features with respect to *x* axis, respectively. This is because for an arbitrary site 1 with the corresponding sliding vector *r* (**Fig. 1g**), we could first anticlockwise rotate the site 1 by 120° to the site 3 and then perform the mirror operation respected to the σ plane to get the site 2, i.e., sites 1 and 2 are symmetric along the *x* axis. Therefore, $\mathcal{D}_{inter}(2) = \sigma \cdot R_{120} \cdot \mathcal{D}_{inter}(1)$, where $\sigma = [(-1/2, \sqrt{3}/2), (\sqrt{3}/2, 1/2)]$, $R_{120} = [(-1/2, \sqrt{3}/2), (\sqrt{3}/2, -1/2)]$, leading to $\mathcal{D}_{inter}^x(2) = \mathcal{D}_{inter}^x(1)$ and $\mathcal{D}_{inter}^y(2) = -\mathcal{D}_{inter}^y(1)$. Different from the in-plane components, the $\mathcal{D}_{inter}^z$ shows a $c_3$ symmetry. The overall next-nearest intralayer DM ($\mathcal{D}_{intra}^{2N}$) is much weaker than that of nearest intralayer DM ($\mathcal{D}_{intra}^{1N}$) [see Supplemental **Fig. S2**], therefore, **Fig. 1h** only presents the nearest intralayer DM vector $\mathcal{D}_{intra}$ ($\mathcal{D}_{intra}^{1N}$). The $\mathcal{D}_{intra}^{1N,x}$ and $\mathcal{D}_{intra}^{1N,y}$ show similar symmetry compared with $\mathcal{D}_{inter}^x$ and $\mathcal{D}_{inter}^y$, but the $c_3$ symmetry for $\mathcal{D}_{intra}^{1N,z}$ is not maintained any more. Surprisingly, although the interlayer distance between two neighboring Cr atoms is significantly larger than that in the intralayer one, the magnitude of $\mathcal{D}_{inter}$ is comparable to that of $\mathcal{D}_{intra}$. This unexpected finding is possibly because during the sliding of top layer atoms, the local symmetry of intralayer is less reduced compared to that of interlayer symmetry. Therefore, solely considering *D*$_{intra}$ but neglecting $D_{inter}$ in twisted moiré magnets may be incorrect to describe the magnetic properties of twisted moiré magnets.

Overall, in BL-CrI$_3$, the $\mathcal{J}_{inter}$ and $\mathcal{J}_{intra}$ are nearly three and over ten times larger than $\mathcal{D}_{inter}$ and $\mathcal{D}_{intra}$, respectively. Furthermore, the calculated perpendicular magnetic anisotropy of Cr atoms is 0.866 meV/Cr. Differing from conventional magnets with homogenous magnetic interactions, the MMEIs in twisted magnets may exhibit a noticeable site-dependent feature, which is further tunable by $\theta$ and $H_{ext}$ [**Eq. (1)**]. Consequently, the competition between $D$ and $J$ varies at different site of the moiré system, i.e., it is possible to generate the noncollinear spin textures at specific local positions in the moiré pattern of tBL-CrI$_3$.

**III. MMEIs-induced Moiré-type SkBs with Non-conserved Helicity.**

Experimentally, the noncollinear spin textures of tBL-CrI$_3$ are discovered when $\theta$ is smaller than 3.00° [19]. Thus, we first take the example of $\theta$~1.41° to discuss the magnetic properties of tBL-CrI$_3$, which can generate a commensurate moiré superlattice with 53,024 atoms and 198,528 nonequivalent MMEI parameters. Starting from **Eq. (1)**, the Landau-Lifshitz-Gilbert (LLG) equation is used to solve the classical atomistic spin problem in tBL-CrI$_3$ (see **Methods**). **Fig. 2a** shows that there are several circular spin textures with two major different sizes distributed in the moiré superlattice. Surprisingly, these two types of circular spin textures are layer-dependent with opposite spin swirling structures, where the large and small textures solely appear in top and bottom layers, respectively. The distributions of normalized magnetic moments along the diameter of circular texture reveal that the moments in the inner and outer regions of these textures are upwards and downwards (see Supplemental **Fig. S3**), respectively, which clearly shows that these microstructures are SkBs [36,37]. The inner and outer diameters of the SkB appeared in top (bottom) layer are ~2.4 and ~9.6 nm, respectively, much smaller than the typically SkBs with the diameters of several hundred nanometers in conventional magnets [38,39].

Interestingly, we discover that these SkBs solely appear surrounding the local AB-stacking sites in the moiré lattice, implying that MMEIs may have strong correlation to the distribution of these SkBs. On one hand, as shown in **Fig. 1e**, the $J_{intra}$ is near constant and is much larger than other magnetic parameters, therefore, $D_{intra}/J_{intra}$ is a relatively small value and negligible. On the other hand, as shown in **Fig. 2b**, the $D_{inter}/J_{inter}$ surrounding the AB-stacking sites [marked by the dashed-white circle (DWC), referred as DWC region hereafter] are much larger than other sites, which could be sufficiently strong to generate noncollinear spin textures (see enlarged DWC structure in Supplemental **Fig. S4**]. Interestingly, if only the shortest-range 1N MMEIs are considered (**Fig. 2c**), instead of SkBs, only collinear FM configurations can be generated (see Supplemental **Fig. S5**), as the $D_{inter}^{1N}/J_{inter}^{1N}$ cannot capture the long-range MMEIs. In other words, the MMEIs at local sites are strongly correlated with the MMEIs at long-range sites.

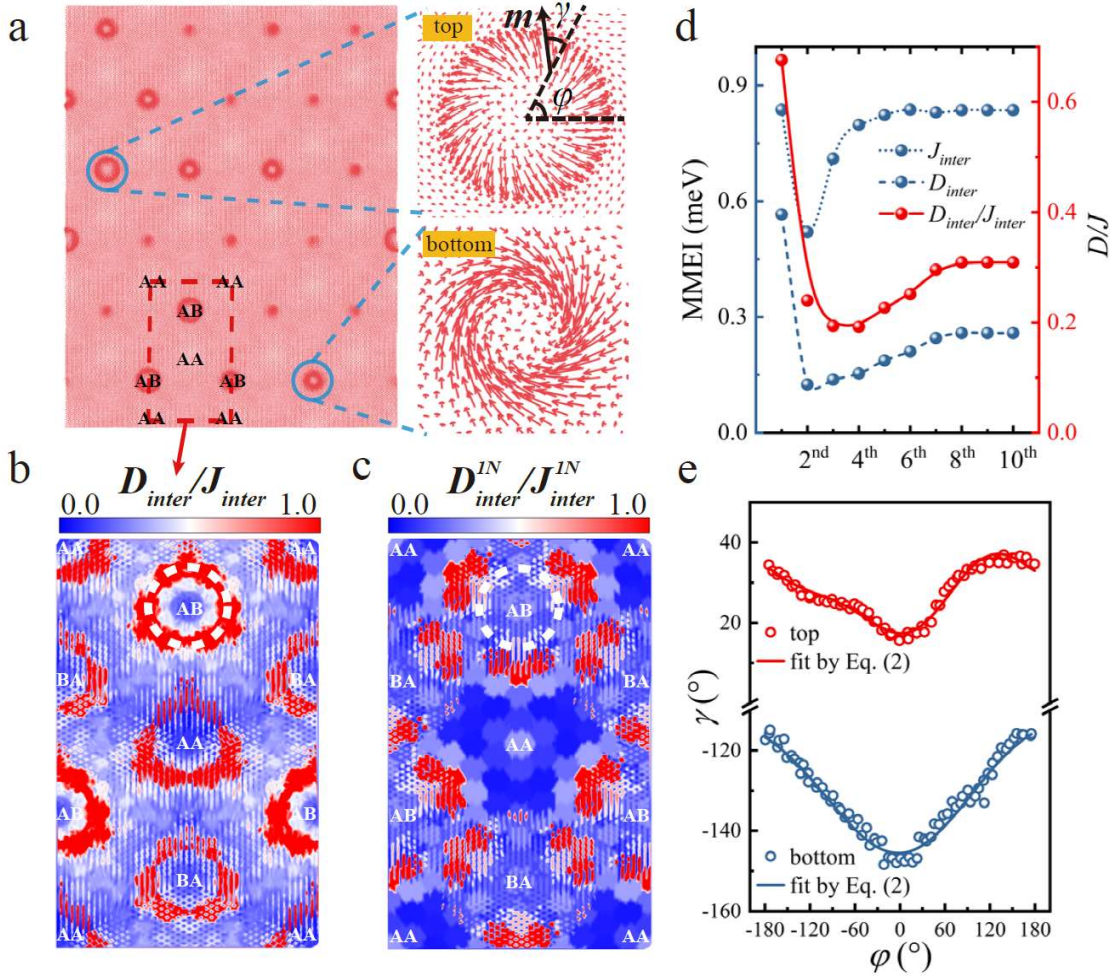

Fig. 2 **MMEIs-induced Moiré-type SkBs at $\theta \sim 1.41°$**. **a**, Spin texture in tBL-CrI$_3$ at $\theta\sim1.41°$. Zoom-in pictures show the SkBs in top and bottom layers, respectively. Rectangular unit cell of moiré superlattice is labeled by dashed-red line. Color mapping for MMEIs ($D_{inter}/J_{inter}$) including **b**, 10N and **c**, 1N magnetic interactions. **d**, Average $J_{inter}$, $D_{inter}$, and $D_{inter}/J_{inter}$ around the DWC region [marked in **b**] as a function of different range of magnetic interactions. **e**, Helicity $\gamma$ as a function of azimuth angle $\varphi$, indicating the $\varphi$-dependent feature.

To further demonstrate the strong correlation between local site and long-range site in MMEIs: (1) we can artificially build a BL-CrI$_3$ with a stacking style similar to the DWC region shown in **Fig. 2b**, i.e., the long-range MMEIs are completely removed. Interestingly, no noncollinear spin textures but only magnetic domains textures can be generated (see Supplemental **Fig. S6**). (2) we plot the average $J_{inter}$, $D_{inter}$ and $D_{inter}/J_{inter}$ which are averaged over the DWC region. As shown in **Fig. 2d**, only if the >8N MMEIs are considered, a converged $D_{inter}/J_{inter}$ can be obtained; only if the > 4N MMEIs are considered, the similar SkBs in **Fig. 2a** can be generated (see Supplemental **Fig. S5**). Therefore, differing from the common belief [22-26], both short-range and long-range MMEIs are critical for describing the local complex spin textures in twisted magnets like tBL-CrI$_3$.

In general, to distinguish the magnetic skyrmions or SkBs, three physical quantities (topological number $N_{sk}$, vorticity $\omega$, helicity $\gamma$) are proposed (see details in Supplemental Note II). For the ordinary Bloch-type and Néel-type skyrmions, they are (-1, 1, ±90°) and (-1, 1, 0 or 180°), respectively [40]. Similarly, for the tBL-CrI$_3$, the $N_{sk}$ and $\omega$ for top and bottom SkBs are also -1 and 1, respectively. Unexpectedly, the $\gamma$ becomes non-conservation, i.e., it changes with the azimuthal angle $\varphi$ (**Fig. 2e**). Remarkably, this non-conserved $\gamma$ has never been discovered or predicted in the existing skyrmions or SkBs. Therefore, we name this new SkB as moiré-type SkBs, reflecting a unique spin texture solely generated by MMEIs. The calculated $\gamma$ for top and bottom SkBs are in ranges of [15°, 37°] and [-148°, -114°], respectively, i.e., they are near opposite (different by $\pi$) with each other, due to the opposite $D$ for distinct layers at the same stacking sites. Interestingly, the $\gamma$ could be fitted by trigonometric function as:

$$\gamma = \sum_n (A_n \sin n\omega\varphi + B_n \cos n\omega\varphi) + C_0 \quad (2)$$

, where $A_1$, $B_1$, $A_2$, $B_2$, $C_0$ and $\omega$ for top (bottom) SkBs are 3.34, -8.11, -1.88, -1.70, 27.18, and 0.018 (0.77, -16.42, -0.10, -1.62, -127.50, and 0.013), respectively. While the $C_0$ approximates the average $\gamma$, the $A_n$, $B_n$, and $\omega$ describe the deviation from averaged $\gamma$ at different $\varphi$. This non-conserved $\gamma$ may induce unconventional physical phenomena, e.g., unconventional skyrmion motion and skyrmion Hall effect.

It is interesting to understand the physical origin of non-conserved $\gamma$. According to Moriya rule [41], $D$ is parallel and perpendicular to the line connecting two magnetic atoms for D$_n$ and C$_{nv}$ point group, respectively. And the corresponding DM energy could be expressed in the Lifshitz invariants form as $\omega_{DM}(D_n) = D^x(L_{zx}^y + L_{yz}^x)$ and $\omega_{DM}(C_{nv}) = D^y(L_{xz}^x + L_{yz}^y)$, respectively. As a result, the Bloch-type skyrmions with helicity of ±90° in D$_n$ point group [42-44] and Néel-type skyrmions with helicity of 0° or 180° in C$_{nv}$ point group [45,46] emerge (see Supplemental **Fig. S7**). As shown in **Fig. 3a**, for C$_n$ point group crystals that can be regarded as subgroup of D$_n$ and C$_{nv}$, there is an angle between DM vector and $r_{AB}$, which is neither 0°, 180° or ±90°. Accordingly, the Lifshitz invariants consist of terms with simultaneous presence of DM interaction with both $x$ and $y$ components, $\omega_{DM}(C_n) = D^x(L_{zx}^y + L_{yz}^x) + D^y(L_{xz}^x + L_{yz}^y)$, and the corresponding helicity is $\gamma = \arctan(-D^x/D^y)$ [47]. Thus, the $\gamma$ is individually determined by the ratio of $D^x$ to $D^y$. For example, the corresponding spin texture with $D^x$=-1 and $D^y$=1 is plotted in bottom of **Fig. 3a**. However, in the tBL-CrI$_3$ system, although the global symmetry belongs to C$_3$ point group, the MMEIs can produce the unique non-conserved $\gamma$ character in the twisted magnets.

To further distinguish the role of MMEIs in contributing to the non-conserved $\gamma$, both the $D_{inter}$ and $D_{intra}$ at the DWC region [marked in **Fig. 2b**] for the top and bottom layers of tBL-CrI$_3$ are plotted in **Fig. 3b**. The local $D_{inter}$ is a near uniform value and gradually winds around the DWC region, resulting in the $\gamma$ near a constant value but deviation from 0°,180° or ±90°. Therefore, the $D_{inter}$ of MMEIs determines the $C_0$ in **Eq. (2)**. While the local $D_{intra}$ is randomly distributed around the DWC region, resulting in the $\gamma$ deviation from a constant value. Therefore, $D_{intra}$ of MMEIs determines the $A_n$, $B_n$, and $\omega$ in **Eq. (2)** and plays a key role in forming non-conserved $\gamma$.

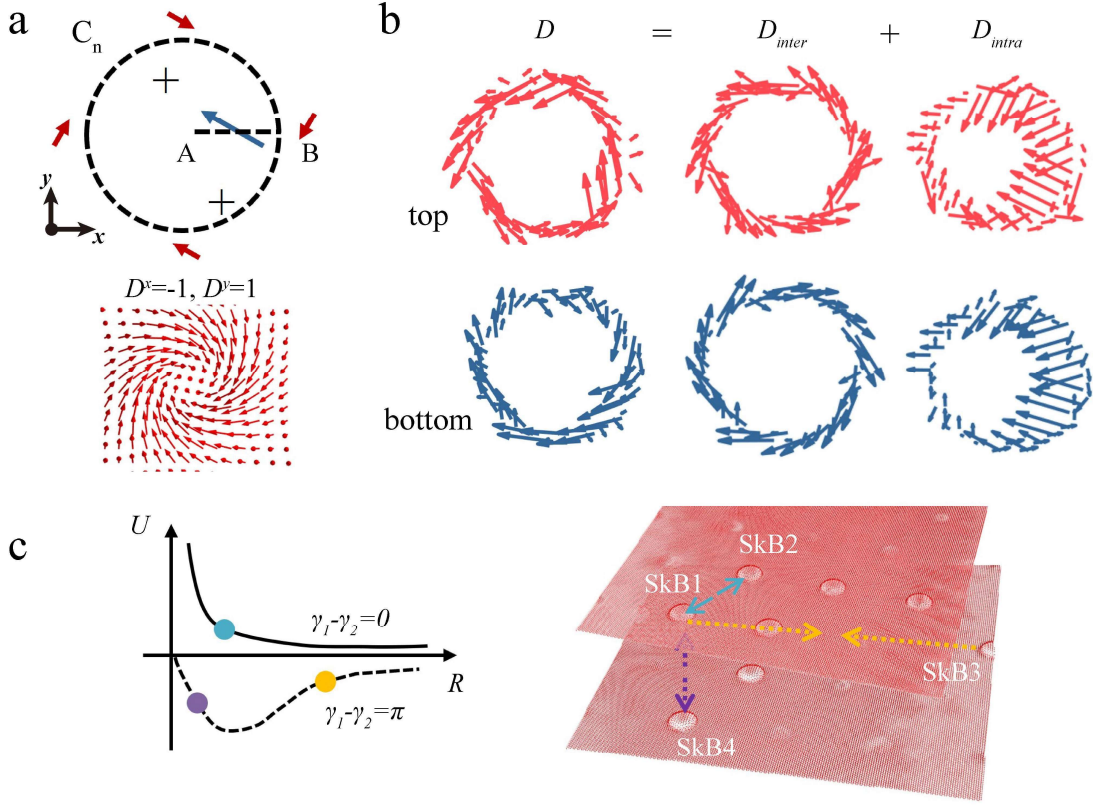

Fig. 3 **Origin of Moiré-type SkBs. a**, Schematic diagram of the DM interactions between two different atoms for $C_n$ point groups. Upper panel: DM vectors (marked with blue arrows) between atoms *A* and *B*, and the corresponding spin of *B* atoms (marked with red arrows). Bottom panel: corresponding spin textures with specific DM parameters. **b**, Local DM distribution around the DWC region [marked in **Fig. 2d**] for top (red arrows) and bottom (blue arrows) layers. **c**, Schematic diagram of the SkB-SkB interactions in the same layer or different layers.

Furthermore, as demonstrated in **Fig. 3c**, it is interesting to understand the basic rules of the interactions of SkBs in the same layer or different layers. For the SkBs located in the same layer, they have the same sign of $\gamma$. Accordingly, the differential of potential *U* as a function of the distance (*R*) between two SkBs in the same layer is always negative (left panel, **Fig. 3c**), giving rise to a repulsion between two neighboring SkBs (e.g., see SkB1 and SkB2 in **Fig. 3c**) [48]. For the SkBs located in the different layers, they have opposite sign of $\gamma$. As shown in left panel of **Fig. 3c**, if the two SkBs are far away from each other, there is an attraction between them (e.g., see SkB1 and SkB3 in **Fig. 3c**); if two SkBs are close to each other, they will be a much stronger repulsion between them (e.g., see SkB1 and SkB4 in **Fig. 3c**). Therefore, the SkBs in top and bottom layers cannot appear at the same stacking site (**Fig. 2a** and Supplemental **Fig. S8**).

Interestingly, the competition of SkB-SkB repulsive and attractive interactions between different layers in the triangular moiré lattice could induce a novel SkB frustrated lattice that was not observed before (see **Fig. S9**).

**VI. Fine Tune of Moiré-type SkBs via θ.**

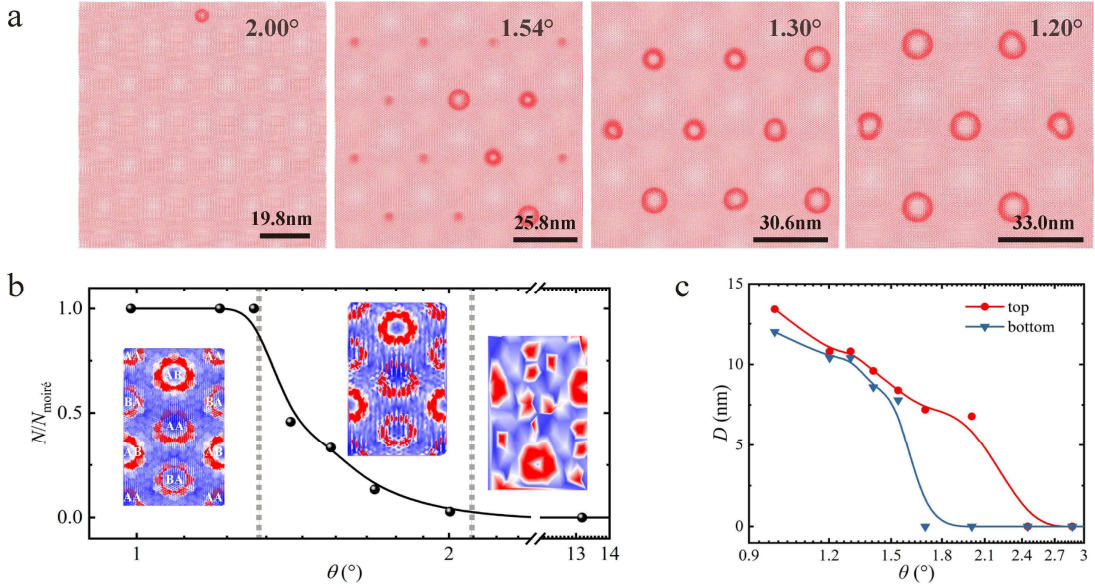

Fig. 4 *θ*-controlled Moiré-type SkBs. **a**, Spin textures for the tBL-CrI$_3$ with different *θ*. Scale bar in each *θ* represents the moiré lattice constant. **b**, Average SkB number for moiré superlattice with different *θ*. Insets: from left to right are three typical color mapping of $D_{inter}/J_{inter}$ at *θ* ~ 1.20°, 2.00° and 13.17°, respectively. **c**, Diameters of SkBs as a function of *θ*.

In a similar way, we have systematically investigated the spin textures in tBL-CrI$_3$ with other commensurate *θ* ranging from 21.78° to 0.99°. In these systems, the number of nonequivalent MMEI parameters vary in a range of 832~403548. Some representative results are shown in **Fig. 4a**. When *θ* > 2.00°, only ordinary moiré patterns with FM spin textures are observed (see supplemental **Fig. S10**). When *θ* = 2.00°, the moiré-type SkBs start to appear, whose number increases with the decrease of *θ*. In particular, when *θ* < 1.30°, each DWC region in the moiré unit cell (marked in **Fig. 2c**) can generate one noncollinear spin texture, which can be SkB or circular domains or antiskyrmions (see Supplemental **Fig. S8a-d**). The ratio of noncollinear spin texture in one moiré unit cell ($N/N_{moiré}$) is summarized in **Fig. 4b**. The trend of $N/N_{moiré}$ as *θ* could be explained by the evolution of $D_{inter}/J_{inter}$. At a large *θ*, there are few magnetic atoms in the moiré unit cell, as a result, the $D_{inter}/J_{inter}$ cannot form clear MMEI patterns to generate noncollinear spin textures (e.g., right inset for *θ* ~ 13.17° in **Fig. 4b**). While as *θ* decreases, the MMEI patterns appear and becomes clear with maximum values at DWC region around AB-stacking positions. As shown in **Fig. 4b**, from left inset with *θ* ~ 1.20° to middle inset with *θ* ~ 2.00°, the smaller the *θ*, the sharper the MMEI pattern. Therefore, $N/N_{moiré}$ increases as function of *θ* decreases from 2.00° to 1.30°. Meanwhile, as the *θ* decreases, the diameters of SkBs monotonously enlarge (**Fig. 4c**), but the domain wall width keeps ~3 nm as $\theta$ decreases (**Fig. S11**). Overall, the non-conserved $\gamma$ is observed in all the SkBs at different $\theta$, showing a similar feature as that in 1.41° (see Supplemental **Fig. S8e**). Therefore, $\theta$-controlled MMEIs plays a key role in determining the population and size of SkB, which may be adopted as a new approach for SkB-based quantum computing and information storage.

**V. Manipulating MMEIs via External Interaction.**

It is interesting to further investigate the role of external substrate on modulating MMEIs. In experiments, to prevent the degradation of tBL-CrI$_3$, they are encapsulated by insulating (hexagonal-BN) films [32]. Importantly, the substrate may further reduce the symmetry of tBL-CrI$_3$. As indicated in **Fig. 5a**, the substrate-induced interfacial DM ($D_{sub}$) may further modulate the MMEIs, changing the $\theta$-dependent MPTs. Given the complexity of the real CrI$_3$/BN interface, we approximate this interfacial DM as a typical value in the range of 0.71~0.87 meV/Cr, based on several typical interfacial configurations [**Fig. S12** and **Table S1**], which is at least five times larger than the maximum $D_{intra}$ in tBL-CrI$_3$. Therefore, the substrate may have a strong influence on the spin textures of tBL-CrI$_3$. In the following, the external interaction term

$$H_{ext} = D_{sub} \sum_{i,j}(\mathbf{S}_i \times \mathbf{S}_j) \quad (3)$$

is used in **Eq. (1)** to simulate the substrate effect.

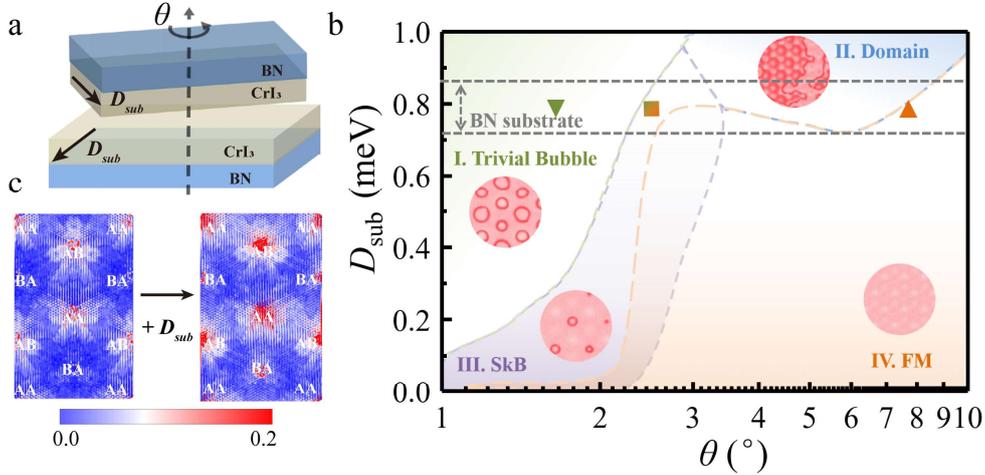

Fig. 5 **Substrate-modulated MMEIs and MPTs. a**, Schematic diagram of tBL-CrI$_3$ encapsulated by BN films, where black arrows indicate the direction of $D_{sub}$. **b**, Magnetic phase diagram of tBL-CrI$_3$ for varying $D_{sub}$ and $\theta$. Insets of **b** show four representative magnetic configurations: (I) trivial bubbles, (II) domains, (III) SkBs, and (IV) FM state. Gray dashed-line region represents the influence of BN substrate. Upper- and lower-triangle represent the regions for experimentally observed FM-domain and FM-AFM-coexisted hexagonal pattern, respectively. Square represents the SkB region that could be detected under the same experimental condition. **c**, Color mapping of $D/J$ before and after the consideration of $D_{sub}$ term induced by BN substrate.

**Fig. 5b** shows the magnetic phase diagram as a function of $D_{sub}$ and $\theta$ (see **Fig. S13** for detailed spin textures). (i) for large commensurate $\theta$, when the $D_{sub}$ is not sufficiently strong, the MMEIs cannot exhibit significant changes, i.e., the systems still exhibit the FM states as that in the free-standing system. For example, the inset IV shows a typical FM spin texture with $D_{sub}$ = 0.00 and $\theta$ = 3.15°. With the influence of $D_{sub}$ is sufficiently strong, the spins will tilt and the tBL-CrI$_3$ exhibits the magnetic domains for large $\theta$. For example, the inset II shows a typical spin texture with $D_{sub}$ = 1.00 meV and $\theta$ = 3.15°. (ii) for relatively small $\theta$, the MMEIs become more important (**Fig. 4b**), as a result, the SkBs appear. For example, the inset III shows a SkB with $D_{sub}$ = 0.00 and $\theta$ = 1.41°. Importantly, with the assistance of strong $D_{sub}$, the $D_{sub}$-modulated MMEIs together with the coulomb repulsion between different SkBs leads to the appearance of trivial bubbles (not SkBs) not only around the AB stacking (i.e., DWC) regions but also around the AA and BA stacking areas, forming the unexpected hexagonal patterns. For example, the inset I shows the trivial bubbles with $D_{sub}$ = 0.80 meV and $\theta$ = 1.20°. This is because, via the $D_{sub}$ term, besides of around AB stacking, the $D/J$ around AA and BA stacking sites can also significantly enlarge, as shown in **Fig. 5c**, which is beneficial for the appearance of new noncollinear spin textures, eventually forming hexagonal patterns. Furthermore, the sizes of these trivial bubbles are >10 nm and the averaged magnetic moment is about 0.7 times $M_s$ of BL-CrI$_3$. Overall, as indicated in **Fig. 5b**, by tuning $D_{sub}$ via different substrates, in principle, different MPTs can be achieved in twisted magnets.

**Discussion.**

*Comparison with experimental observations.* The conditions for the experimentally observed FM region and FM-AFM-coexisted region with hexagonal spin textures in tBL-CrI$_3$ are labeled as upper- and lower-triangle in **Fig. 5b**, respectively. Interestingly, these two spin textures well fall into IV and I regions, perfectly agreeing with our calculations. In particular, the transition region (marked as square) between $\theta$=2°~3° is very similar to the experimentally observed critical point with the coexistence of FM and FM-AFM phases [19]. Therefore, it is reasonable to believe that the near hexagonal periodic noncollinear spin textures with nonzero remanent magnetic signal observed in experiments at small $\theta$ could be essentially the trivial magnetic bubbles. Interestingly, in this transition region, the unconventional moiré-type SkBs can also exist, which is very likely to be detected in the current experimental samples. In practice, the circular dichroism in resonant elastic x-ray scattering may be useful to visualize the non-conserved helicity of the SkBs [49].

*Comparison with the existing models.* All the existing models reported in the literature can only partially describe the magnetic interactions in the tBL-CrI$_3$ by artificially treating the MMEIs with different approximations, e.g., employing homogenous $J_{intra}$ and $D_{intra}$, layer-averaged $J_{inter}$, or neglecting $D_{inter}$ [23-26]. Importantly, based on the accurate consideration of MMEIs, all the spin textures reported in the previous models can also be reproduced by our moiré spin Hamiltonian [**Eq.(1)**], except for the stacking domain-wall magnon states [22]. Therefore, **Eq. (1)** also has high compatibility with the existing simplified spin models. Most importantly, our model can perfectly explain the current experimental observations, confirming its validity.

**Outlook and Conclusion.**

As one of the most fundamental physical quantities in twisted magnets, the MMEIs have not been well understood yet, due to its extreme complexity. Using the developed microscopic moiré spin Hamiltonian, for the first time, the MMEIs can be effectively captured and considered via a novel sliding-mapping approach. Interestingly, a new moiré-type SkB state with non-conserved helicity has been discovered, reflecting the unique role of MMEIs in generating unconventional spin textures. The creation and annihilation of these SkBs along with their sizes can be finely achieved by tuning $\theta$. Importantly, combining the MMEIs and substrate effect, we can well explain the experimentally observed complex spin textures along with its physical origin, solving the outstanding debates between prior theories and experiments. Besides of substrates, it is expected that the MMEIs may also been tuned by external electric fields, strain fields, or charge doping, providing exotic ways to tune the nonlinear spin textures and MPTs in twisted magnets. We note that the anisotropic Kitaev [50] and other small high-order magnetic interactions[51] may also have some influence on the MMEIs. In addition, the noncollinear magnetism may lead the emergence of intrinsic electric polarization by the KNB model [52] and induce electric polarization in the tBL-$CrI_3$ system [30], tuning the MMEIs for device applications. In general, our developed spin Hamiltonian may be widely adopted to predict new spin textures or understand the spin-related physical phenomena in other twisted magnets.

**Methods**

**DFT calculations**

The DFT calculations are performed using the Vienna ab initio simulations package (VASP) [53]. The core electrons were treated using the projector augmented wave (PAW) method and the correlation potential was treated using the generalized gradient approximation (GGA) with the Perdew-Burke-Ernzerhof (PBE) functional [54]. The optB86b vdW correction was employed in all calculations [55,56]. The Dudarev's method [57] with an effective Hubbard $U$ = 3.0 eV was used to correct the correlation effects caused by the partially occupied Cr-3$d$ orbitals. A 300-eV energy cutoff and a 9 × 9 × 1 Γ-centered $k$-mesh were adopted to perform calculations. To avoid the artificial interaction caused by periodic boundary condition, a vacuum of 15 Å was added. The in-plane lattice constant of BL-$CrI_3$ was fixed to 6.93 Å for different stacking models. All the atoms were fully relaxed along the $z$-direction in bilayer $CrI_3$ system until the force and total energy was less than 0.01 eV/Å and $10^{-5}$ eV. In our simulation, we construct the $CrI_3$/BN heterojunction with 1×1 $CrI_3$ lattice and 3×3 BN.

**Mapping of magnetic interactions**

To remove the artificial strain effect in the twisted bilayer systems, only commensurate $\theta$ was considered. The atomic positions in tBL-CrI$_3$ moiré crystal were mapping from the different stacking bilayer structures. The magnetic interactions, including exchange interaction $J$ and $D$, were calculated based on the Green's function method as implemented in the TB2J package [58]. The single-ion anisotropy was calculated using the force theorem [59,60]. The tBL-CrI$_3$ in our calculation with largest twisted angle 21.78° and smallest twisted angle 0.99° contain 28 (832) and 13468 (403548) magnetic Cr atoms (MMEI parameters), and the corresponding lattice constants were 18.34 Å and 402.12 Å, respectively. To alleviate the consumption of computation resources, we used the sliding-mapping method to obtain the magnetic interactions in tBL-CrI$_3$, which is based on identifying the relative position of Cr-Cr pair in tBL-CrI$_3$ ($d_{mx}, d_{my}$) and BL-CrI$_3$ ($d_x, d_y$) that are match with each other. That is to say, ($d_{mx}, d_{my}$)=($d_x, d_y$), we could get the $D$ and $J$ in tBL-CrI$_3$ simultaneously. The details of mapping of MMEIs can be found in Supplemental Note I.

**Atomic spin simulations**

The spin dynamics simulations were performed using the VAMPIRE package [61]. The time evolution of the magnetization was described by the Landau-Lifshitz-Gilbert (LLG) equation and was integrated numerically using the Heun numerical scheme. Here, to have a more convenient and higher efficient calculation on the tBL-CrI$_3$ spin textures, the least integral multiple of the rectangular moiré lattices with the size larger than 100 nm×100 nm was adopted for each system, which contains over 96,000 magnetic atoms. The periodic boundary conditions were used to perform calculations. The initial spin textures were selected as the randomly distributed spin textures. All simulations were performed using the zero-field cool mode which the calculation temperature starts at 30 K and decreases in steps of 1 K to near 0K. Therefore, the spin textures observed in the tBL-CrI$_3$ are very likely to be the ground-state configurations. For each temperature step, $3\times10^6$ calculation steps were performed at a fixed time step of 0.1 fs.

**Acknowledgements**

The authors acknowledge the support from National Key Research and Development Program of China (Grant No. 2022YFA1402400), National Natural Science Foundation of China (Grant No. 12088101) and NSAF Grant U2230402. All the calculations were performed at Tianhe2-JK at CSRC.

**Author contributions**

B.H. supervised and led the project. B.Y. and B.H. directed the project. B.Y., Y.L., and B.H. prepared the manuscript. All authors discussed the results and contributed to the manuscript.

**Competing interests**